\documentclass[a4paper]{jpconf}
\usepackage{graphicx}
\usepackage{amsmath}
\begin{document}
\title{The Bethe-Salpeter approach to bound states: from  Euclidean to Minkowski space.}

\author{
A. Castro $^a$, E. Ydrefors$^a$, W. de Paula$^a$, T. Frederico$^a$,
J.H. de Alvarenga Nogueira$^{a,b}$, P. Maris$^c$}
\address{
  $^a$Instituto Tecnol\'ogico da Aeron\'autica, DCTA, 12.228-900 S\~ao Jos\'e dos Campos, SP,  Brazil\\
  $^b$Universit\`a di Roma “La Sapienza”, INFN, Sezione di Roma, P.le A. Moro 5, 00187 Roma, Italy \\
  $^c$Department of Physics and Astronomy, Iowa State University, Ames, IA 50011, USA
}

\begin{abstract}
The challenge to obtain from the Euclidean Bethe--Salpeter amplitude
the amplitude in Minkowski is solved by resorting to un-Wick rotating
the Euclidean homogeneous integral equation.  The results obtained
with this new practical method for the amputated Bethe--Salpeter
amplitude for a two-boson bound state reveals a rich analytic
structure of this amplitude, which can be traced back to the Minkowski
space Bethe--Salpeter equation using the Nakanishi integral
representation.  The method can be extended to small rotation angles
bringing the Euclidean solution closer to the Minkowski one and could
allow in principle the extraction of the longitudinal parton density
functions and momentum distribution amplitude, for example.
\end{abstract}

\section{Introduction}
Techniques to solve the Bethe--Salpeter Equation (BSE) in Minkowski
space have been developed for bound state of bosons
\cite{KusPRD95,KusPRD97,Sauli:2001mb,FrePR12,PimentelFBS16} and
fermions \cite{CarEPJA10,dPaPRD16,dPaEPJC17}, at the expense of being
algebraically quite involved, either by use of the Nakanishi integral
representation (NIR) \cite{Nak69} or by direct integration.  On the
other hand, calculations done in Euclidean space after performing the
Wick rotation of the BSE are conceptually
straigthforward~\cite{Maris:1997tm,Maris:2003vk}, but it is nontrivial
to obtain structure observables that are defined on the light-front,
such as e.g. parton distributions, from Euclidean solutions.  It is
desirable to be able to undo the Wick rotation and obtain the
Minkowski space solutions from the Euclidean solutions, such that one
could extract Minkowski space observables.  The first steps in this
direction are provided in this contribution.
 
Our goal here is to present solutions of the BSE for two-bosons close
to the Minkowski space, by introducing a rotation into the complex
plane of $k_0\to k_0\exp(\imath\theta)$, where $\theta=\pi/2$ is the
standard Wick-rotation associated with the Euclidean space
formulation, while the Minkowski space formulation corresponds to
$\theta=0$.  We present solutions of the BSE for small angles and show
that the rich analytic structure is accessible numerically by such
technique~\cite{inprep}.  The branch-points obtained from the integral
representation of the vertex function are exhibited by our accurate
numerical solutions, for the two-boson bound state in ladder
approximation.  We present an initial study with angles small as
$\theta=\pi/128\approx 1.4^\circ$, where we also explore different
masses $\mu$ of the exchanged boson, as well as binding energies.

\section{Two-body BSE in Euclidean space  \label{BSEucl}}
A two-body bound state with total four-momentum $p$ with $p^2 = -M^2$
can be described by the amputated Bethe--Salpeter vertex function
$\Gamma$, which is a solution of the two-body bound state equation
\begin{eqnarray}
  \Gamma(k^2, k\cdot p; p^2=-M^2) & = &
  \int\!\!\frac{d^4k'}{(2\pi)^4} \; K(k-k'; p) \;
  \Delta(\tfrac{p}{2}  + k') \; \Gamma(k'^2, k'\cdot p; p^2 = -M^2) \; \Delta(\tfrac{p}{2}  - k') \,. \nonumber\\
\end{eqnarray}
Here, $K$ is the two-body scattering kernel, and $\Delta$ are the
(dressed) propagators for the constituent particles.  In ladder
truncation, the kernel reduces to $\frac{g^2}{(k-k')^2 + \mu^2}$.
Using bare propagators in the rest frame with 3-dimensional spherical
coordinates, $p = (iM, \vec{p}=0)$, we have for the scalar bound state
in the Euclidean metric
\begin{eqnarray}\label{bseuc}
  \Gamma(k_0, k_v; i\,M)
   & = & - \frac{m^2 \, \alpha}{\pi^2}
  \int_{-\infty}^{+\infty}\!\!dk'_0  \int_{0}^{+\infty}\!\!dk'_v \; \frac{k'_v}{k_v}\;
  \ln{\left(\frac{(k_0-k'_0)^2 + (  k_v-k_v')^2 + \mu^2 }{(k_0-k'_0)^2 + ( k_v+k'_v)^2 + \mu^2 } \right)} 
  \nonumber \\
  && {} \times
  \frac{\Gamma(k'_0, k'_v; i\,M)}{[ (\frac{i}{2} M + k'_0)^2 + {k'_v}^2 + m^2 ] \; [ (\frac{i}{2} M - k'_0)^2 + {k'_v}^2 + m^2 ] } \; ,
\end{eqnarray}
where $k_v\equiv|\vec k_v\,|$ and $k'_v\equiv|\vec k'_v\,|$,
and $\alpha = g^2 / (16 \pi m^2)$.  The corresponding canonical
normalization condition \cite{IZ} can be written as
\begin{eqnarray}
 N^2 & = & \frac{1}{4\pi^3} \int_{-\infty}^{+\infty} \!\!dk_0  \int_{0}^{+\infty} \!\!dk_v \;
 \frac{k^2_v \; \left[\Gamma(k_0,  k_v; i\,M)\right]^2}
      {\left[(\frac{i}{2} M + k_0)^2 +  k^2_v + m^2 \right]^2 \; \left[(\frac{i}{2} M - k_0)^2 +  k^2_v + m^2 \right] } \;
\end{eqnarray}
such that $\Gamma(k_0, k_v; i\,M) / N$ is the properly normalized
amputated Bethe--Salpeter vertex.

Although BSE is usually solved in the rest frame of the bound state,
it should be noted that the (amputated) vertex $\Gamma(k^2, k\cdot p;
p^2=-M^2)$ is a function of the Lorentz scalar variables $k^2$ and
$k\cdot p$ at fixed $p^2$.  As long as the truncation of the BSE does
not break Lorentz invariance (including any regularization schemes of
divergences), the obtained solution is frame independent, and
therefore does not need to be boosted for e.g. form factor
calculations.  Indeed, it has been demonstrated that meson
observables, including form factors, calculated in the ladder
truncation, are indeed
frame-independent~\cite{Maris:2005tt,Bhagwat:2006pu}.

\section{Un-Wick rotation towards Minkowski space \label{BSWick}}
\begin{figure}[thb] 
  \includegraphics[width=0.45\textwidth]{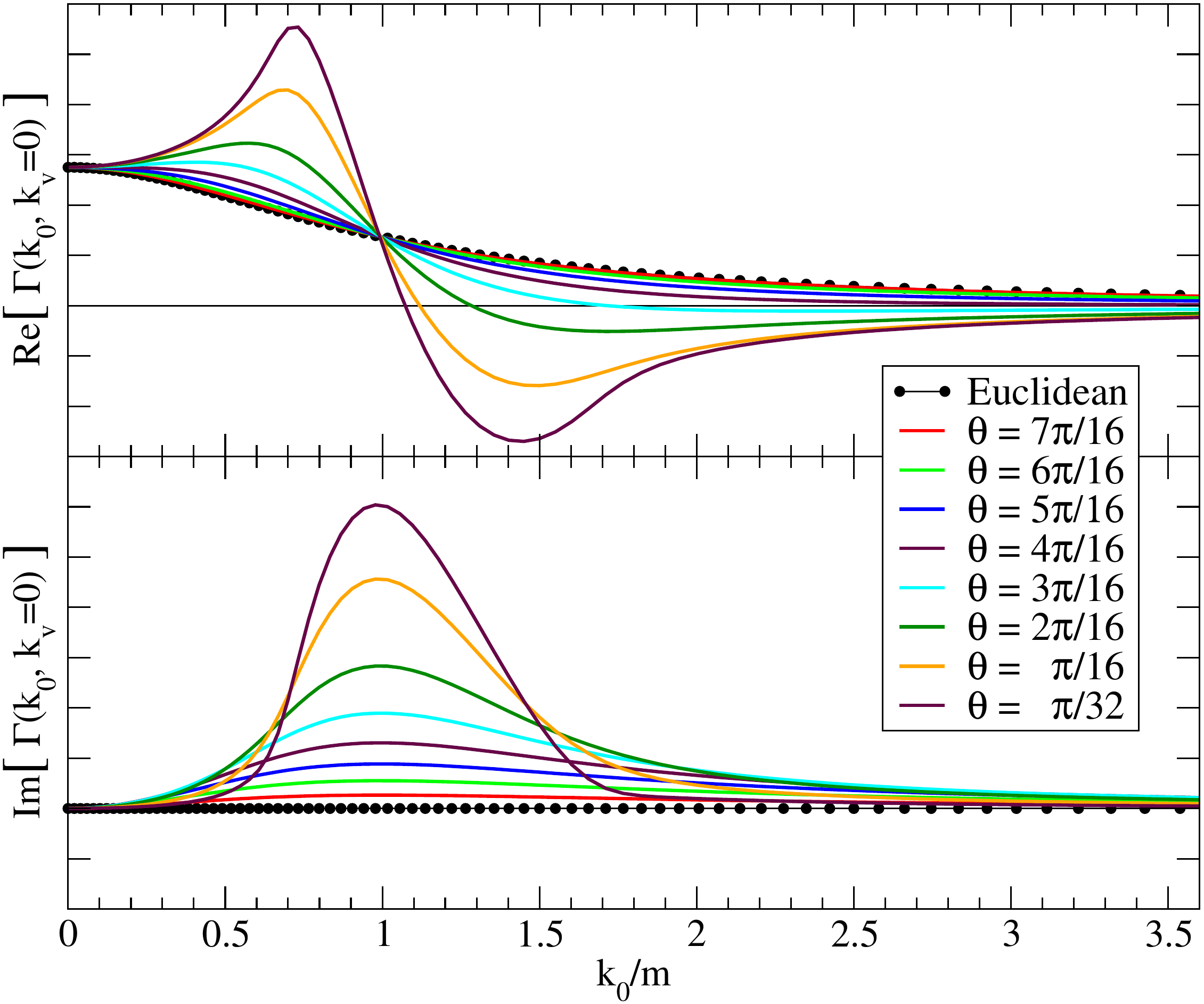}
  \qquad
  \includegraphics[width=0.45\textwidth]{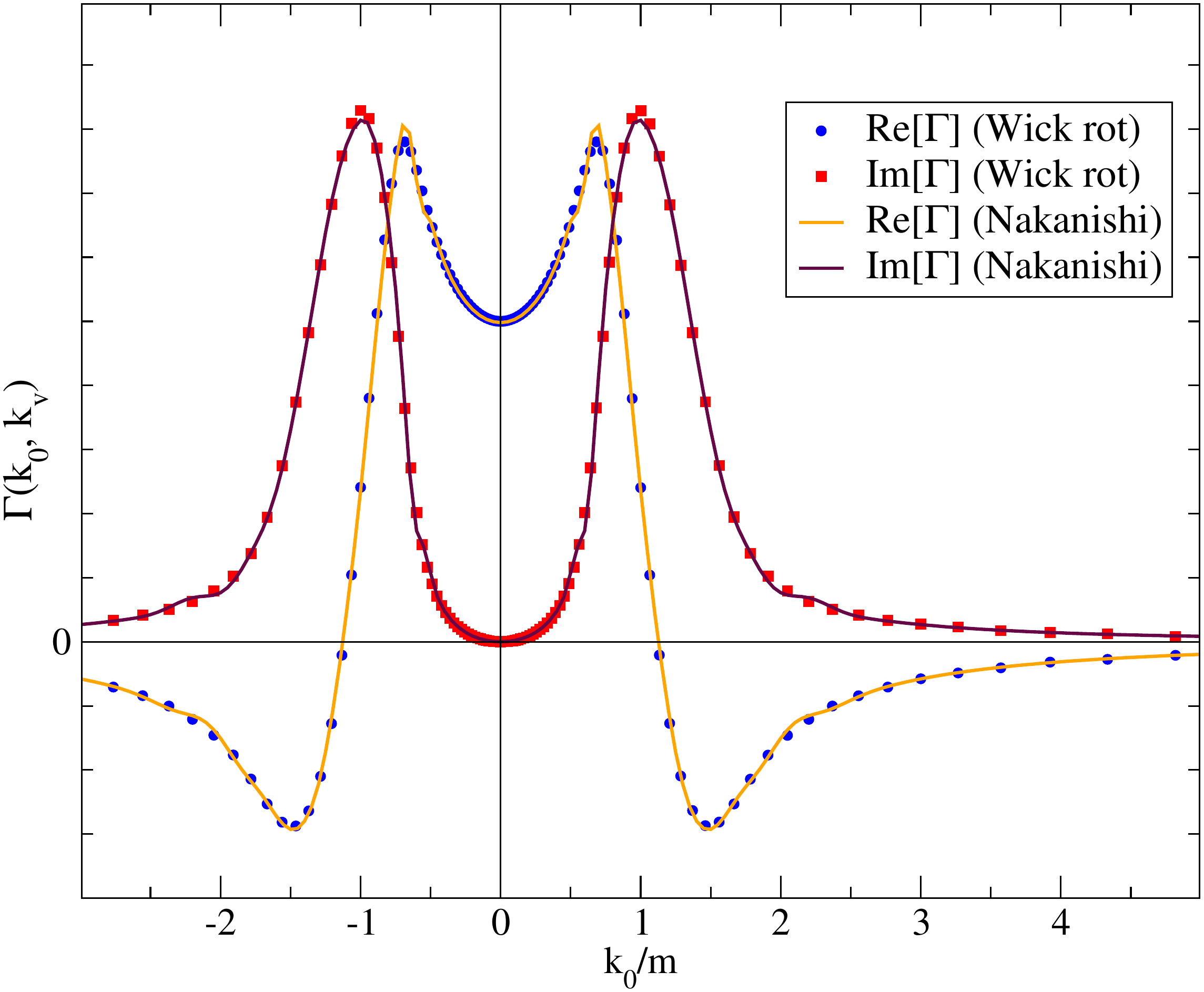}
  \caption{ \label{Fig:theta_Naka}
    Numerical solutions for $\Gamma(k_0, k_v)$ for $\mu/m=0.2$ and $M/m=1.0$
    in arbitrary units.
    Left: Solutions of the un-Wick rotated Euclidean BSE for a range of $\theta$;
    Right: Comparison of the un-Wick rotated Euclidean BSE and
    the NIR at $\theta=\pi/16\approx11^\circ$.
  }
\end{figure}

Starting with the Euclidean space ladder BSE the rest frame,
Eq.~(\ref{bseuc}), we can make a change of variables
$k_0 \to k_0 \, {\rm e}^{i\delta}$ and $k'_0 \to k'_0 \, {\rm e}^{i\delta}$
with the rotation angle $ \delta=\theta-\pi/2 $, where $\theta$ is the
rotation angle with respect to the usual Minkowski definition of the
zero component of the momentum.  Thus the un-Wick rotated BSE becomes
\begin{eqnarray}
  \Gamma({\rm e}^{i\delta}k_0, k_v; i\,M) 
  & = & - \frac{m^2 \, \alpha\,{\rm e}^{i\delta}}{\pi^2}
  \int_{-\infty}^{+\infty} \!\!dk'_0 \int_{0}^{+\infty} \!\!dk'_v \; \frac{ k'_v}{k_v} \; 
  \ln{\left(  \frac{{\rm e}^{2 i\delta}(k_0-k'_0)^2 + (k_v-k'_v)^2 + \mu^2 }{{\rm e}^{2 i\delta}(k_0-k'_0)^2 + (k_v+k'_v)^2 + \mu^2 } \right)}\; 
  \nonumber \\
  && {} \times
\frac{\Gamma({\rm e}^{i\delta}k'_0, k'_v; i\,M)}
     {[(\frac{i}{2} M + {\rm e}^{i\delta}k'_0)^2 + k'^2_v + m^2 ]\;[(\frac{i}{2} M - {\rm e}^{i\delta}k'_0)^2 + k'^2_v + m^2 ]}\,,         
\end{eqnarray}
which can be solved numerically, e.g. by iteration.  In particular,
starting from the Euclidean solution ($\delta = 0$), one can increase
$\delta$ in small steps, and at each step use the solution at the
previous step as the initial guess for solving the BSE iteratively,
as is illustrated in the left panel of Fig.~\ref{Fig:theta_Naka}.
Of course, as one approaches the Minkowski axis ($\delta \to \pi/2$,
or equivalently, $\theta \to 0$), the numerical challenges in order to
obtain a stable solution increase.  Although we cannot solve the BSE
at $\theta=0$ exactly, we may be able to extrapolate to $\theta=0$.

\section{Two-body BSE  in Minkowski space}
Alternatively, one can use the NIR to solve the BSE in Minkowski space
formulation.  Following Ref.~\cite{KusPRD95}, we make use of the
uniqueness assumption of the Nakanishi weight function in the
non-perturbative domain.  We have to remind that for the
Bethe--Salpeter amplitude itself, the uniqueness assumption can be
overcome, using the method of Light-Front projection \cite{KarEPJA06},
followed by the application of the inverse generalized Stieltjes
transform \cite{Carbonell:2017kqa}.

The integral representation of the vertex function $\Gamma$ is
\begin{eqnarray}\label{vnaka}
  \Gamma(k^2, k\cdot p; p^2 = M^2) &=&
  \int^{+1}_{-1}dz\int^\infty_{\gamma_{\text {min}}} d\gamma \,
  \frac{g_\Gamma(\gamma,\,z)}{\gamma+m^2-\frac{p^2}{4}-k^2-k\cdot p\,z-\imath\epsilon}\;.
\end{eqnarray}
A task we have to undertake is to determine the minimum value of
$\gamma$ by checking for the adequacy of the solution in the form
above for the BSE, which can in principle depend on $z$
\cite{Wanders57}.  After introducing the one-boson exchange kernel in
the BSE, and using uniqueness, we find that \cite{KusPRD95}
\begin{eqnarray*}
  g_\Gamma(\gamma,\,z) = \frac{g^2}{(4\pi)^2}
  \int_{-1}^{+1} \!\! dz' \int_0^1 \!\! d\alpha_2 \int_0^1 \!\! d\alpha_3\,
  \frac{(1-\alpha_3)}{(1+z') \, \bar{s} }
  \theta(\bar{s}) \,
  \theta(1-\bar{\alpha}) \,
  \theta(\bar{\alpha})\,
  \theta(\gamma_0 - \gamma_{\min})  \,
  \frac{\partial g_\Gamma(\gamma',\,z')}{\partial \gamma' } \bigg |_{\gamma'=\gamma_0} \, ,
\end{eqnarray*}
where
\begin{eqnarray*} 
  \bar{s} &=&\frac{(1+z)+2z'(\alpha_2+\alpha_3)+\alpha_3(1-z)}{1+z'}
  \nonumber \\
  \bar\alpha &=&\frac{\alpha_2(1-z')+(z'-z)(1-\alpha_3)}{1+z'}
  \nonumber \\
  \gamma_0 &=&\frac{\alpha_3(1-\alpha_3)\gamma - (1-\alpha_3)^2 \,
    \left(m^2 + (z^2-1)\,\frac{p^2}{4}\right)-\alpha_3 \, \mu^2 }{\bar{s}}
  \nonumber \\
  \gamma_{\text {min}} &=&\mu \,\left(2 \sqrt{m^2 + (z^2-1) \,\frac{p^2}{4}} + \mu\right) \; .
\end{eqnarray*}
After a redefinition of the parameter $\gamma$, the integral equation
for $g_\Gamma(\gamma,\,z)$ can be solved numerically using basis
expansion (see e.g. \cite{FrePRD14}), and from that the observables
like parton distributions can be calculated.

As a consistency check, we can also apply the (un-)Wick rotation to
the NIR, and calculate the vertex function
$\Gamma({\rm e}^{i\theta}k_0, k_v; M)$ from $g_\Gamma(\gamma,\,z)$.
We do indeed find good agreement between the un-Wick rotated solution
of the Euclidean BSE and the solution at arbitrary angles $\theta$
from the NIR, as can be seen in the right panel of Fig.~\ref{Fig:theta_Naka}.

\section{Analytic structure of the Bethe--Salpeter amplitude}
Our numerical solutions shown in the left panel of
Fig.~\ref{Fig:theta_Naka} strongly suggest the existence of
singularities in the amputated vertex function $\Gamma$.  A detailed
analysis of the NIR, Eq.~(\ref{vnaka}), shows that there are indeed
branch-points in the amputated vertex function, located for $z=\pm 1$
at
\begin{eqnarray}
  \gamma_{\text{min}} + m^2-\frac{p^2}{4} - k^2 \pm k\cdot p
  &=& ( m +\mu)^2 - \frac{p^2}{4}- k^2\pm k\cdot p \; = \;  0 \, ,
\end{eqnarray}
which in the rest frame gives the branch-points at
\begin{eqnarray} \label{bp}
  |k_0| & = &  k_0^{\pm}  \quad \equiv \quad \sqrt{(m+\mu)^2+k_v^2} \pm \frac{M}{2} \,.
\end{eqnarray}
The positive and negative branch-points in $k_0$ closest to the origin
are separated by $2\sqrt{(m+\mu)^2+k_v^2} - M$, which allows the
rotation of the arguments of the vertex function in the complex $k_0$
plane without crossing singularities.  This non-analytic behavior of
the vertex function at these branch-points should be corroborated by
the numerical results found for $\Gamma(k; p)$ in the $k_0$ plane.

\begin{figure}[thb] 
  \includegraphics[width=0.45\textwidth]{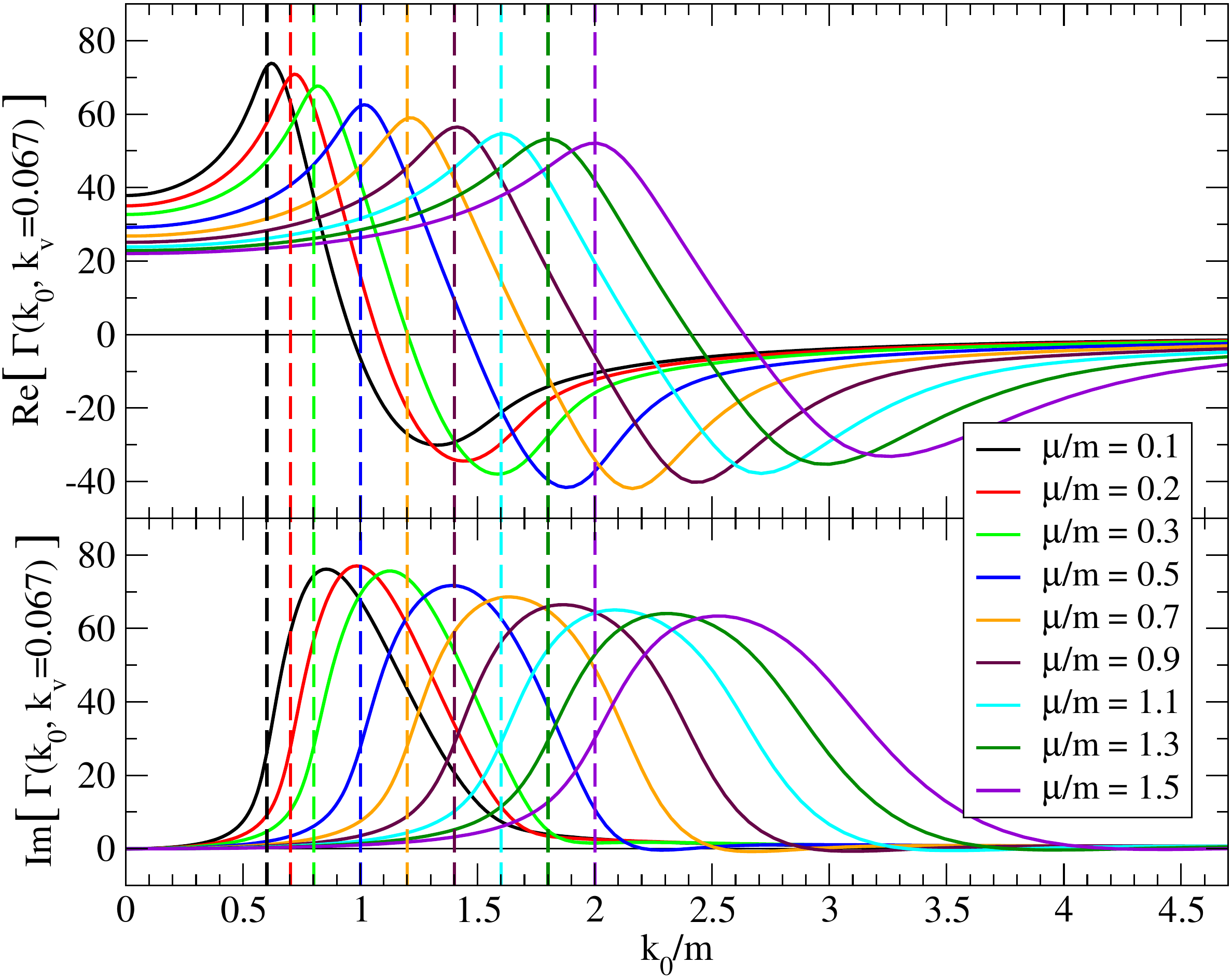}
  \qquad
  \includegraphics[width=0.45\textwidth]{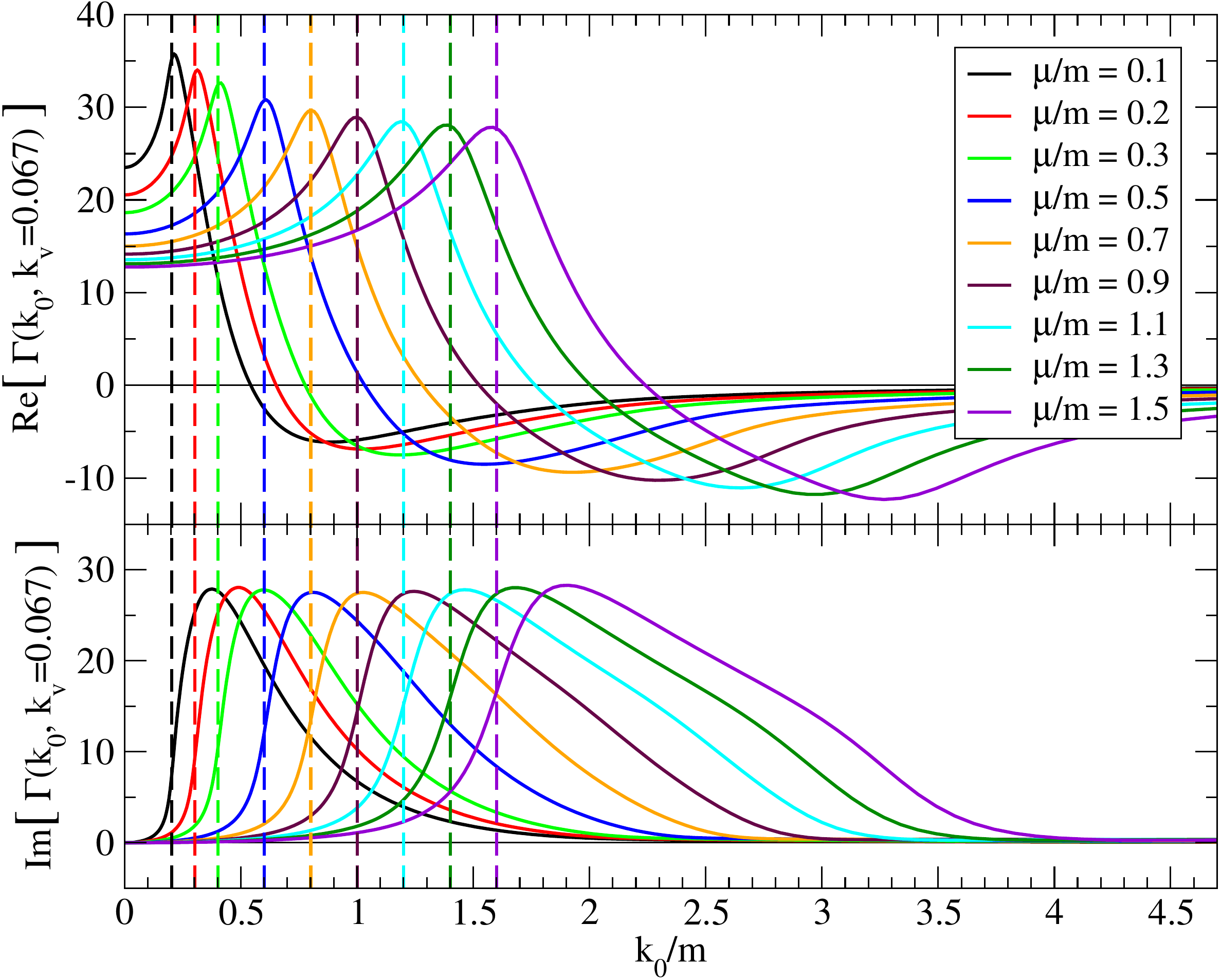}
  \caption{ \label{Fig:BSA_mu}
    Numerical solutions for $\Gamma(k_0, k_v)$ at
    $\theta=\pi/32\approx5.6^\circ$ for a range of exchange masses
    $\mu$.
    Left: for moderate binding, $M/m = 1.0$.
    Right: for weak binding, $M/m = 1.8$.
    The dashed vertical lines indicate the location of the first
    branch-point in $\Gamma(k_0, k_v)$.
  }
\end{figure}

In Fig.~\ref{Fig:BSA_mu}, we present our results with
$0.1\le\mu/m\le1.5$ at an angle $\theta=\pi/32\approx5.6^\circ$ for
two different bound state masses: $M/m=1$, corresponding to moderate
binding, and $M/m=1.8$, corresponding to weak binding.  The vertical
bars show the location of the branch-points $k_0^-$, see
Eq.~(\ref{bp}), in the limit $\theta \to 0$.  For
$\theta=\pi/32\approx5.6^\circ$ the real part of $\Gamma(k_0, k_v)$
has a peak for $|k_0| \approx k^-_0$.  Furthermore, the imaginary part
of $\Gamma(k_0, k_v)$ is (almost) zero for $|k_0| < k^-_0$, but rises
sharply near for $|k_0| \approx k^-_0$.  At fixed binding energy,
these peaks are more pronounced as the mass of the exchange particle
decreases to zero.

\begin{figure}[thb] 
  \includegraphics[width=0.44\textwidth]{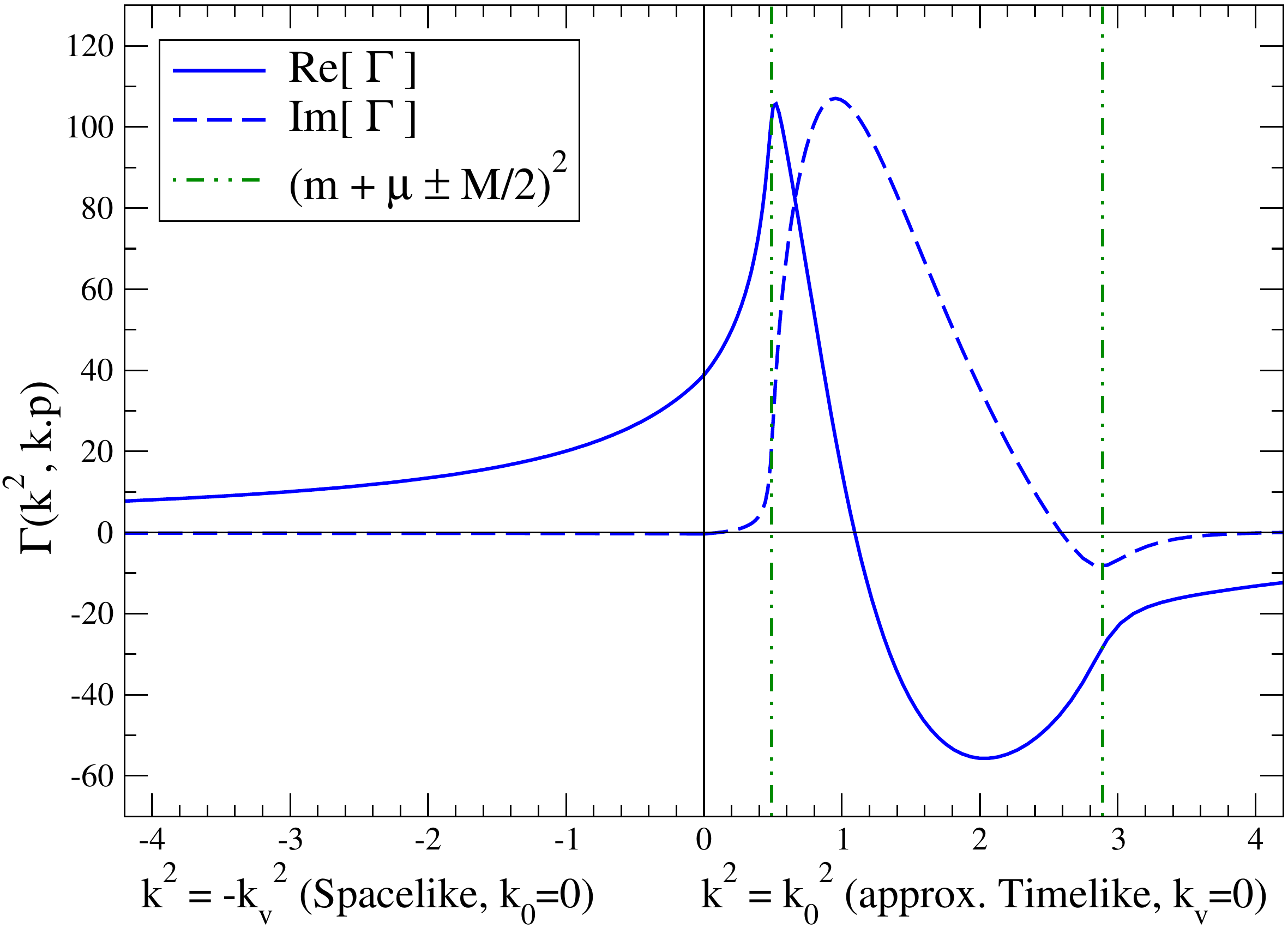}
  \qquad
  \includegraphics[width=0.49\textwidth]{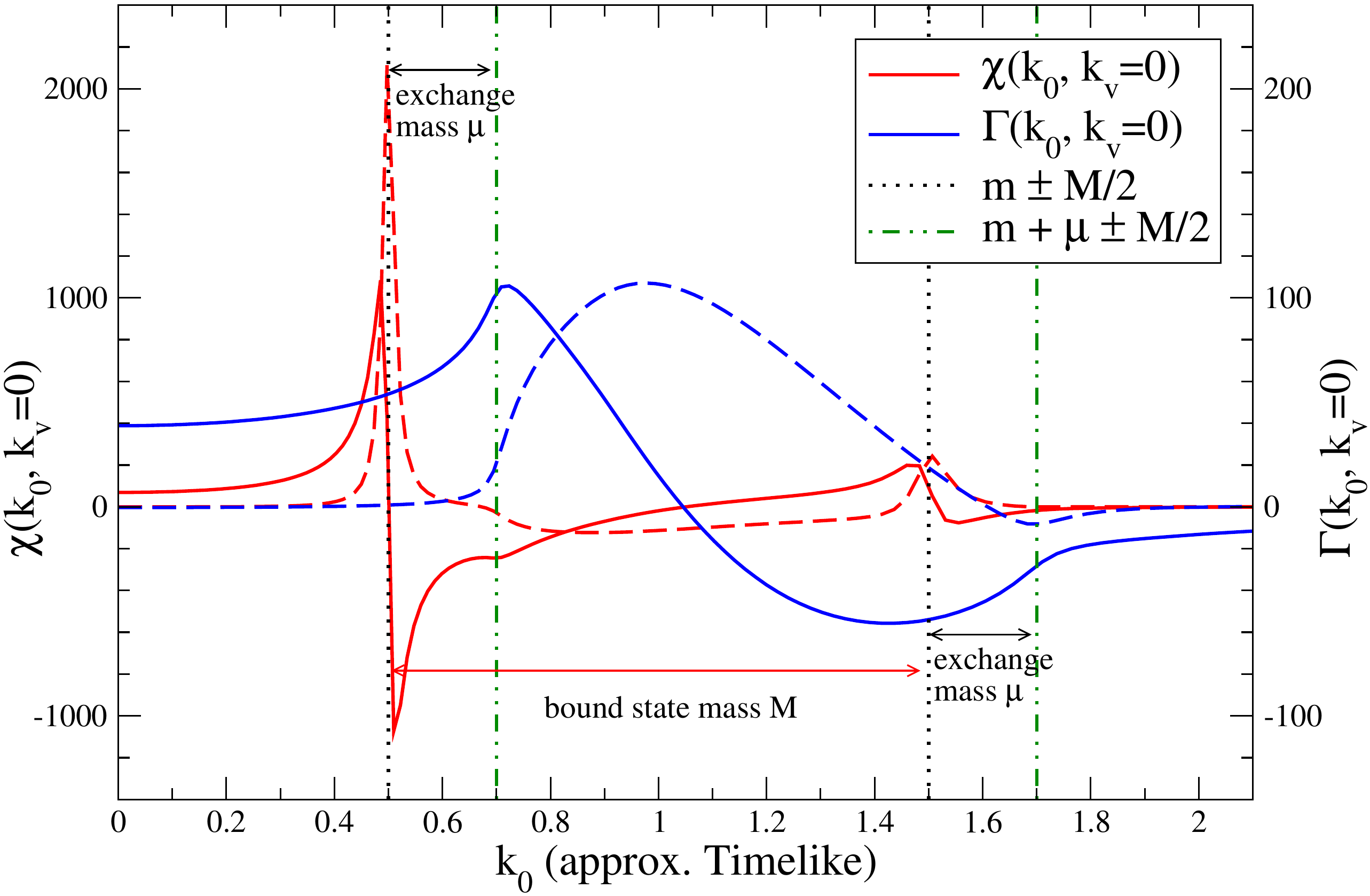}
  \caption{
    Numerical solution of the BSE for $\mu/m = 0.2$ and $M/m = 1.0$
    at $\theta=\pi/128\approx1.4^\circ$.
    Left: $\Gamma(k^2, k\cdot p)$ as function of $k^2$ in both the
    spacelike and (approximate) timelike region;
    Right: results for both $\Gamma$ (blue) and $\chi = \Delta \Gamma
    \Delta$ (red) as function of $k_0$.
    \label{Fig:TimelikeBSA} }
\end{figure}

As one decreases the angle $\theta$ to approach the Minkowski axis,
both the peak in the real part of $\Gamma(k_0, k_v)$ and the sharp
rise in the imaginary part of $\Gamma(k_0, k_v)$ become more
pronounced, see the left panel of Fig.~\ref{Fig:TimelikeBSA},
suggesting that this is indeed a branch-point.  These high-precision
numerical calculations also confirm that there are no singularities
closer to $k_0=0$ in $\Gamma(k_0, k_v)$ than those at $|k_0| = k^-_0$.  Furthermore, the
kink in both the real and the imaginary parts of $\Gamma(k_0, k_v)$
indicate the location of the non-analytic points at $|k_0| = k_0^+$.

Finally, in the right panel of Fig.~\ref{Fig:TimelikeBSA} we show
the Bethe--Salpeter amplitude with the external propagator legs
\begin{eqnarray}
  \chi(k_0, k_v; p) &=&
  \Delta(\tfrac{p}{2}  + k') \; \Gamma(k_0, k_v; p) \; \Delta(\tfrac{p}{2}  - k) \,,
\end{eqnarray}
in addition to $\Gamma(k_0, k_v; p)$.  Here we clearly see that the
analytic structure of $\chi(k_0, k_v; p)$ is dominated by the poles in
the constituent propagators $\Delta$; the additional non-analytic
structure at $|k_0| = k_0^-$ is reduced to relatively minor kinks 
in the real and imaginary parts of $\chi(k_0, k_v; p)$ at
$|k_0|=k_0^-$, and the non-analyticity of $\chi(k_0, k_v; p)$ at
$|k_0|=k_0^+$ is not even visible in this plot.

\section{Concluding remarks}
In this work we present a method to solve the BSE for two-bosons close
the timelike axis in Minkowski space.  To this end we perform an
un-Wick rotation of the Euclidean BSE into the $k_0$ complex plane.
Our solutions of this un-Wick rotated BSE are in good agreement with
solutions obtained by solving the BSE in Minkowski space using the
Nakanishi Integral Representation and a posteriori rotation into the
complex plane.  The numerical solutions suggest the existence of
branch-points as one approaches the timelike region.  Indeed, a
detailed analysis of the Minkowski space Bethe--Salpeter equation
using the Nakanishi Integral Representation reveals a rich analytic
structure of the Bethe--Salpeter amplitude.

In conclusion, the un-Wick rotation captures the main physics of the
vertex function, and it can be a valuable tool in the study of the
Bethe--Salpeter amplitude close to the timelike region.  We expect
that this method can be useful to obtain structure observables that
are defined on the light-front, such as e.g. parton distributions,
and to further explore the phenomenology of strongly relativistic
bound state systems.

%
\ack
  We thank FAPESP Thematic grants no. 13/26258-4 and no. 17/05660-0.
  PM thanks the Visiting Researcher Fellowship from FAPESP, grant no. 2017/19371-0;
  EY thanks FAPESP grant no. 2016/25143-7;
  JHAN thanks FAPESP grant no. 2014/19094-8;
  TF thanks Conselho Nacional de Desenvolvimento Cient\'ifico e
  Tecnol\'ogico (Brazil) and Project INCT-FNA Proc. No. 464898/2014-5.
  This study was financed in part by CAPES - Finance Code 001.

\hspace{1cm}


\section*{References}
\begin{thebibliography}{99}

\bibitem{KusPRD95} K. Kusaka and A. G. Williams, Phys. Rev. D {\bf 51} (1995) 7026.

\bibitem{KusPRD97} K. Kusaka, K. Simpson and A. G. Williams, Phys. Rev. D {\bf 56} (1997) 5071.

\bibitem{Sauli:2001mb}
  V.~Sauli and J.~Adam,
  Nucl.\ Phys.\ A {\bf 689} (2001) 467.
  
\bibitem{FrePR12} T. Frederico, G. Salm\`e and M. Viviani, Phy. Rev. D {\bf 85} (2012) 036009.
  
\bibitem{PimentelFBS16} R. Pimentel, W. de Paula,  Few Body Syst. {\bf 57} (2016) 7, 491.

\bibitem{CarEPJA10} J. Carbonell, V.A. Karmanov, Eur. Phys. J. A {\bf 46} (2010) 387.

\bibitem{dPaPRD16} 	
W. de Paula, T. Frederico, G. Salm\`e, M. Viviani, Phys. Rev. D {\bf 94} (2016) 071901.

\bibitem{dPaEPJC17} 	
W. de Paula, T. Frederico, G. Salm\`e, M. Viviani, R. Pimentel, Eur. Phys. J. C {\bf 77} (2017) 11, 764. 

\bibitem{Nak69} N. Nakanishi, Suppl. Prog. Theor. Phys. {\bf 43} (1969) 1.

\bibitem{Maris:1997tm}
  P.~Maris and C.~D.~Roberts,
  Phys.\ Rev.\ C {\bf 56} (1997) 3369.

\bibitem{Maris:2003vk}
  P.~Maris and C.~D.~Roberts,
  Int.\ J.\ Mod.\ Phys.\ E {\bf 12} (2003) 297.
  
\bibitem{inprep} A. Castro {\it et al}, in preparation.

\bibitem{IZ} Claude Itzykson and Jean-Bernard Zuber,  Quantum Field Theory (McGraw-Hill, 1985)

\bibitem{Maris:2005tt}
  P.~Maris and P.~C.~Tandy,
  Nucl.\ Phys.\ Proc.\ Suppl.\  {\bf 161} (2006) 136.

\bibitem{Bhagwat:2006pu}
  M.~S.~Bhagwat and P.~Maris,
  Phys.\ Rev.\ C {\bf 77} (2008) 025203.

\bibitem{KarEPJA06} V. A. Karmanov and J. Carbonell, Eur. Phys. J. A {\bf 27} (2006) 1.

\bibitem{Carbonell:2017kqa} 
  J.~Carbonell, T.~Frederico and V.~A.~Karmanov,
  Phys.\ Lett.\ B {\bf 769}, 418 (2017).

\bibitem{Wanders57} G. Wanders, Helvetica Physica Acta {\bf 30}  (1957) 417.

\bibitem{FrePRD14} T. Frederico, G. Salm\`e and M. Viviani, Phy. Rev. D {\bf 89} (2014) 016010.

\end{thebibliography}
\end{document}